\documentclass[a4paper,11pt]{article}

\usepackage{pos}
\usepackage{epsfig}
\usepackage{graphicx}
\usepackage{amsmath}
\usepackage{bm}
\usepackage{bbm}
\usepackage{multirow}

\title{Born--Oppenheimer EFT: a unified description of ordinary and exotic quarkonia}
\ShortTitle{Born--Oppenheimer EFT}

\author*[a]{Antonio Vairo}

\affiliation[a]{Technical University of Munich,\\
TUM School of Natural Sciences, Physics Department,\\   
James-Franck-Str.~1, 85748 Garching, Germany}

\emailAdd{antonio.vairo@tum.de}

\abstract{We show how the Born--Oppenheimer effective field theory (BOEFT) provides a unified description of ordinary and
  exotic quarkonia grounded on the non-relativistic expansions of QCD and supplemented with lattice QCD inputs.
We apply BOEFT to tetraquarks, pentaquarks, quarkonium hybrids and to assess threshold effects in the quarkonium spectrum.
}

\FullConference{The 21st International Conference on Hadron Spectroscopy and Structure (HADRON2025)\\
 27 - 31 March, 2025\\
Osaka University, Japan\\}

\begin{document}
\maketitle

\section{Introduction: XYZ states}
QCD allows for more color singlet combinations of quarks and gluons than conventional hadrons, i.e. mesons and baryons \cite{Gell-Mann:1964ewy}.
These exotic combinations have remained for a long time elusive until at the beginning of the century B-factories started to
detect non-conventional states in the charmonium and bottomonium spectrum generically dubbed {\it XYZ states} (for a review see~\cite{Brambilla:2019esw}).
Some of them contain hidden charm or bottom but are charged, $T_{c\bar{c}}^+(4050)$, ..., $T_{b\bar{b}1}^+(10650)$, ...~.
These and other properties allow to identify some of the non-conventional states as tetraquarks made of two heavy and two light quarks.
Other states, with much less certainty, could be quarkonium hybrid candidates.
More recently pentaquark states made of two heavy and three light quarks have also been detected.
Since the discovery of the $\chi_{c1}(3872)$ in 2003 by Belle, 45 possibly non-conventional hadrons have been discovered in the charmonium and bottomonium spectrum 
(see \href{https://qwg.ph.nat.tum.de/exoticshub}{https://qwg.ph.nat.tum.de/exoticshub}).

\section{BOEFT}
Quarkonia and quarkonium exotica are bound states made of pairs of heavy quarks (we denote with $Q$ the heavy quark and with $\bar{Q}$ the heavy antiquark).
The quarks being heavy guarantees that the hierarchy of non-relativistic energy scales $m_Q \gg p \sim 1/r \sim m_Q v \gg E \sim m_Q v^2$,  
where $m_Q$ is the heavy quark mass, $p$ the relative momentum, $r$ the relative distance, $E$ the binding energy and $v$ the heavy quark relative velocity,
is, at least parametrically, fulfilled.
The hierarchy of energy scales calls for a hierarchy of effective field theories (EFTs)~\cite{Brambilla:2004jw}.
The ultimate EFT is the {\it Born--Oppenheimer EFT} (BOEFT)~\cite{Berwein:2015vca,Oncala:2017hop,Brambilla:2017uyf,Soto:2020xpm,Berwein:2024ztx}
that reduces to potential NRQCD for heavy quarkonia~\cite{Brambilla:1999xf}.
At first order, BOEFT reproduces the Born--Oppenheimer approximation: the heavy quarks move adiabatically in the presence of the light degrees of freedom,
whose effect is encoded in a suitable set of potentials that depend on the distance $\bm{r}$ of the heavy quarks.

$QQ$ ($Q\bar{Q}$) states are classified according to the symmetry group $D_{\infty h}$, whose representations are labeled $\Lambda^\sigma_\eta$: 
if $\bm{k}$ is the angular momentum of the light degrees of freedom, $|\bm{r}\cdot\bm{k}| = \Lambda = 0,1,2,... \equiv \Sigma, \Pi, \Delta, ...$,   
$\eta$ is the P(C) eigenvalue ($g\equiv 1$ and $u\equiv -1$), and $\sigma$ is the reflection eigenvalue (only for $\Sigma$).
The numbers $\Lambda$, $\sigma$ and $\eta$ are called {\it Born--Oppenheimer (BO) quantum numbers}.
Higher states for a given irreducible representation are labeled by primes, e.g. $\Sigma_g$, $\Sigma_g^\prime$,~...~.
For $r \to 0$, the symmetry group becomes $O(3)$ ($\times C$).  
Hence several $\Lambda_\eta^\sigma$ representations reduce to one single $k^{P(C)}$ representation \cite{Foster:1998wu}, see table~\ref{tab1}. 

\begin{table}[h!]
\makebox[0truecm]{\phantom b}\put(70,0){
\begin{tabular}{|c|c|}  \hline
  $k^{PC}$ & BO quantum \#  \\
\hline\hline
$0^{++}$ & $\Sigma_g^+$ \\
$0^{+-}$ & $\Sigma_u^+$ \\
$0^{-+}$ & $\Sigma_u^-$ \\
$1^{+-}$ & $\{\Sigma_u^-,\Pi_u\}$ \\
$1^{--}$ & $\{\Sigma_g^+,\Pi_g\}$ \\
$2^{--}$ & $\{\Sigma_g^-,\Pi_g,\Delta_g\}$ \\
\hline
\end{tabular}}
\put(250,7){
\begin{tabular}{|c|c|}  \hline
  $k^{P}$ & BO quantum \#  \\
\hline\hline
$0^{+}$ & $\Sigma_g^+$ \\
$0^{-}$ & $\Sigma_u^-$ \\
$1^{+}$ & $\{\Sigma_g^-,\Pi_g\}$ \\
$1^{-}$ & $\{\Sigma_u^+,\Pi_u\}$ \\
$2^{-}$ & $\{\Sigma_u^-,\Pi_u,\Delta_u\}$ \\
\hline
\end{tabular}}
\caption{\label{tab1} $Q\bar{Q}$ (left) and $QQ$ (right) $\Lambda_\eta^\sigma$ multiplets and their corresponding $k^{PC}$ and $k^P$ multiplets.}
\end{table}

At leading order in the expansion in $v$ the BOEFT Lagrangian can be written as 
\begin{align}
L_{\mathrm {BOEFT}}=&\int d^3{\bm R}\int d^3{\bm r} \, \sum_{\lambda\lambda^{\prime}}\mathrm{Tr}\Bigg\{{\Psi}^{\dagger}_{\kappa\lambda}(\bm{r},\,\bm{R},\,t) \bigg[i\partial_t\,\delta_{\lambda\lambda'}-
V_{\kappa, \lambda\lambda^\prime }(\bm{r})   \nonumber\\
&\hspace{2.5 cm}+ \sum_\alpha P^{\alpha\dag}_{\kappa\lambda}\left(\theta, \varphi\right)\frac{\bm{\nabla}^2_r}{m_Q}P^{\alpha}_{\kappa\lambda^{\prime}}\left(\theta, \varphi\right)\bigg]
{\Psi}_{\kappa\lambda^{\prime}}(\bm{r},\,\bm{R},\,t)\Bigg\} + \dots  ~,
\label{lag_BOEFT}
\end{align}
where $\kappa = \{ k^{P(C)},{\rm flavor} \}$ are the light degrees of freedom quantum numbers, 
$\lambda = \pm \Lambda$, and the fields $\Psi_{\kappa\lambda}$ identify the states, with $P^\alpha_{\kappa\lambda}$ suitable projectors.
The potentials $V_{\kappa, \lambda\lambda^\prime }(\bm{r})$ can be expanded in $1/m_Q$:
\begin{align*}
&V_{\kappa\lambda\lambda^{\prime}}(\bm{r}) = V_{\kappa\lambda\lambda'}^{(0)}(r)  + \frac{V_{\kappa\lambda\lambda'}^{(1)}(\bm{r})}{m_Q} +  ...~,
\end{align*}
with $V_{\kappa\lambda\lambda'}^{(0)}(r)$ the static potential (in the following written as $V_{\Lambda_\eta^\sigma}(r)$)
and $V_{\kappa\lambda\lambda'}^{(1)}(\bm{r})$ including the leading spin term.

The potentials appearing in the BOEFT Lagrangian include the contributions to the dynamics from the non-perturbative soft degrees of freedom.
Their form can be determined by lattice QCD computations.
In general, potentials with the same BOEFT quantum numbers mix.

\begin{figure}
\begin{center}
{\epsfxsize=9truecm \epsfbox{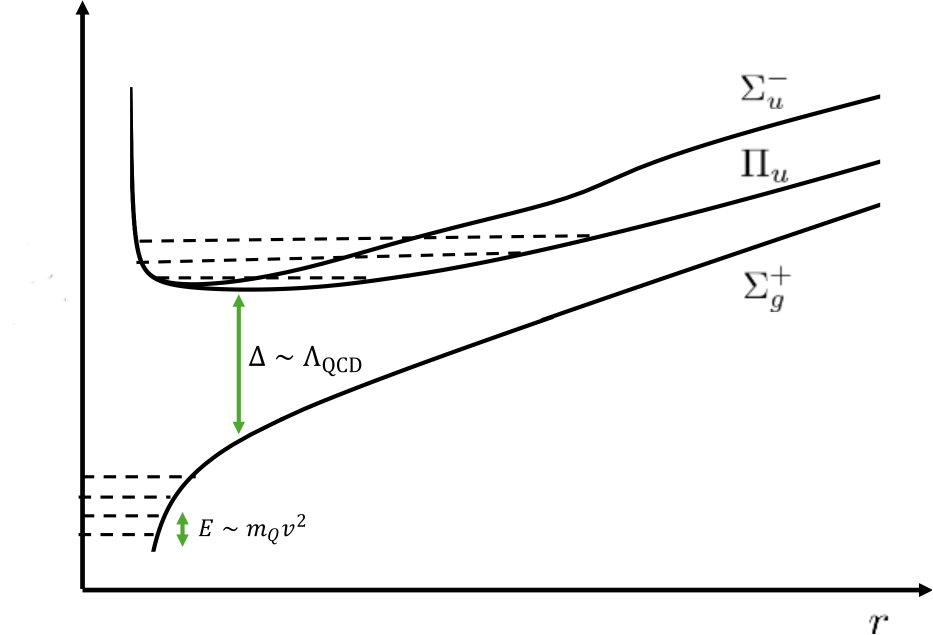}}
\caption{\label{fig2} Isospin $I=0$ $\Sigma_g^+$, $\Pi_u$ and $\Sigma_u^-$ potentials at short distance $r$ of the static sources.
Energy gaps between states are also displayed.} 
\end{center}
\end{figure}

{\bf Mixing at short distance.}
Because of the degeneracy pattern shown in table~\ref{tab1}, at short distance also BOEFT potentials with different Born--Oppenheimer quantum numbers, but with the same $O(3)$ quantum numbers, mix.
The situation is schematically illustrated for the isospin $I=0$ $\Sigma_g^+$, $\Pi_u$ and $\Sigma_u^-$ potentials in figure~\ref{fig2}.
If we consider potentials between a heavy $Q$ and $\bar{Q}$, the $\Sigma_g^+$ potential is the quarkonium potential.
The excited potentials support quarkonium hybrids or isospin $I=0$ tetraquarks.
The gap of order $\Lambda_{\rm QCD}$ between the $\Sigma_g^+$ potential and the excited potentials guarantees that there is no significant mixing between quarkonium and excited states at short distance,
so that in fact a Cornell-like potential describes well quarkonium below threshold.  
At very short distance, excited potentials between $Q$ and $\bar{Q}$ pairs behave like a repulsive Coulomb potential between color octet sources.
At short distance, they group into degenerate multiplets with the same {\it gluelump} mass in the case of hybrids or {\it adjoint meson} mass in the case of $Q\bar{Q}q\bar{q}$ tetraquarks,
reflecting the restored $O(3)$ symmetry of table~\ref{tab1}.
Similarly they group into degenerate multiplets with the same {\it triplet meson} mass in the case of low-lying $QQ\bar{q}\bar{q}$ tetraquarks
and {\it adjoint baryon} mass in the case of  $Q\bar{Q}qqq$ pentaquarks.

The equations of motion that follow from \eqref{lag_BOEFT} are coupled Schr\"odinger equations reflecting the mixing pattern~\cite{Berwein:2015vca,Berwein:2024ztx}.
Accounting for the short-distance mixing/degeneracy of the $\Pi_u$ and $\Sigma_u^-$ potentials leads to the coupled radial Schr\"odinger equations
for the states (hybrids, tetraquarks) supported by the potentials $\Pi_u$ and $\Sigma_u^-$:
\begin{align}
  &\Bigg[-\frac{1}{m_Qr^2}\,\partial_rr^2\partial_r+\frac{1}{m_Qr^2}\begin{pmatrix} l(l+1)+2 & -2\sqrt{l(l+1)} \\ -2\sqrt{l(l+1)} & l(l+1) \end{pmatrix} \nonumber\\
    &\hspace{4.5 cm}+\begin{pmatrix} V_{\Sigma_{u}^{-}}(r) & 0 \\ 0 & V_{\Pi_{u}}(r) \end{pmatrix}\Bigg]\hspace{-4pt}\begin{pmatrix} \psi_{\Sigma, \sigma_P}^{(N)} \\
    \psi_{\Pi, \sigma_P}^{(N)}\end{pmatrix}=\mathcal{E}_N\begin{pmatrix} \psi_{\Sigma, \sigma_P}^{(N)} \\
  \psi_{\Pi, \sigma_P}^{(N)}\end{pmatrix},
\label{Sch_short1}
\end{align}
for $k=1$ and parity $\sigma_P$ ($l(l+1)$ is the eigenvalue of $\bm{L}^2 = (\bm{k} + \bm{L}_Q)^2$, $\bm{L}_Q$ is the orbital angular momentum of the heavy quark pair),
and to the Schr\"odinger equation
\begin{align}
  \left[-\frac{1}{m_Qr^2}\,\partial_r\,r^2\,\partial_r+\frac{l(l+1)}{m_Qr^2}+V_{\Pi_{u}}(r)\right]\psi_{\Pi, -\sigma_P}^{(N)}=\mathcal{E}_N\,\psi_{\Pi, -\sigma_P}^{(N)},
\label{Sch_short2}
\end{align}
for the opposite parity $-\sigma_P$.
Note the {\it parity doubling} phenomenon, i.e. the lifting of the parity degeneracy in the spectrum.
The solutions of the Schr\"odinger equations provide the radial parts of the state wavefunctions and the binding energies $\mathcal{E}_N$.

\begin{table}[th!]
\small{\renewcommand{\arraystretch}{1}
\begin{minipage}{.49\linewidth}
\begin{tabular}{|c|c|c|}  \hline
  \multirow{2}{*}{$\begin{array}{c} k_{\bar{q}}^{P}\otimes k_q^{P}\end{array} $ } & \multirow{2}{*}{ $\begin{array}{c}k^{PC}\end{array} $ } &  \multirow{2}{*}{$\begin{array}{c} \text{BO quantum \#}\\ \end{array} $ }\\& & \\
\hline\hline
$(1/2)^-\otimes(1/2)^+ $         & $0^{-+}$ & $\Sigma_u^-$ \\
& $1^{--}$ & $\Sigma_g^+,\,\Pi_g$ \\
\hline
\end{tabular}
\end{minipage}
\begin{minipage}{.49\linewidth}
\begin{tabular}{|c|c|c|}  \hline
\multirow{2}{*}{$\begin{array}{c}k_{\bar{q}}^{P}\otimes k_{\bar{q}}^{P}\end{array} $ } & \multirow{2}{*}{ $\begin{array}{c}k^{P}\end{array} $ } &  \multirow{2}{*}{$\begin{array}{c} \text{BO quantum \#}\\ \end{array} $ }\\& & \\
\hline\hline
$(1/2)^-\otimes(1/2)^- $         & $0^{+}$ & $\Sigma_g^+$ \\
                                & $1^{+}$ & $\Sigma_g^-,\,\Pi_g$ \\ \hline
\end{tabular}
\end{minipage}
\caption{\label{tab2} The $k^{PC}$ quantum numbers of the light quark-antiquark $\left(\bar{q} q\right)$ pair (left) and $k^{P}$ quantum numbers of the light antiquark-antiquark $\left(\bar{q} \bar{q}\right)$ pair (right)
  for the lightest light quark states contributing to S-wave plus S-wave thresholds of static heavy mesons-antimeson and static heavy meson-meson pairs, respectively.
  The light quark states have quantum numbers $k_{q}^P$ and $k_{\bar{q}}^P$, $k_q$ ($k_{\bar{q}}$) being the angular momentum of the light (anti)quark.
  The Born--Oppenheimer quantum numbers corresponding to the $k^{P(C)}$ quantum numbers are shown in the third columns.}}
\end{table}

\begin{figure}
\makebox[0cm]{\phantom b}\put(10,0){\epsfxsize=7truecm \epsffile{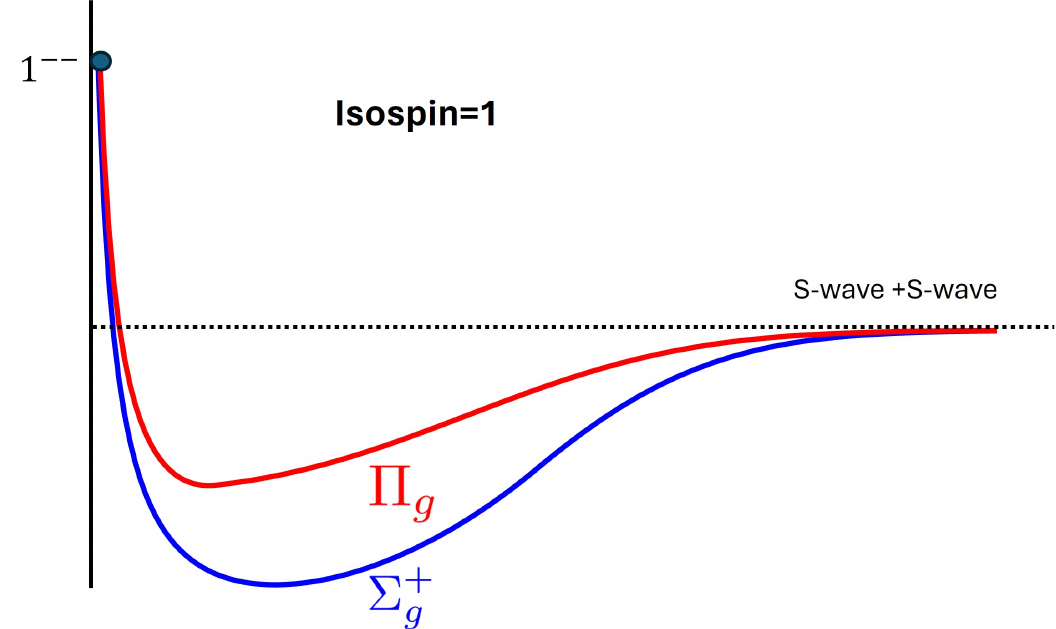}}
\put(210,0){\epsfxsize=7truecm \epsffile{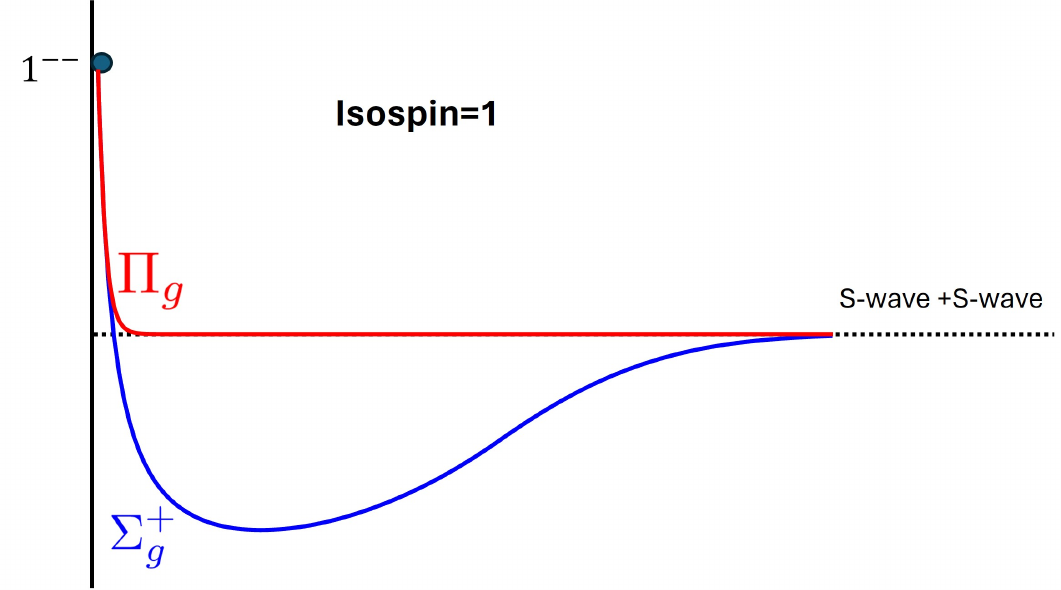}}
\put(90,-150){\epsfxsize=7truecm \epsffile{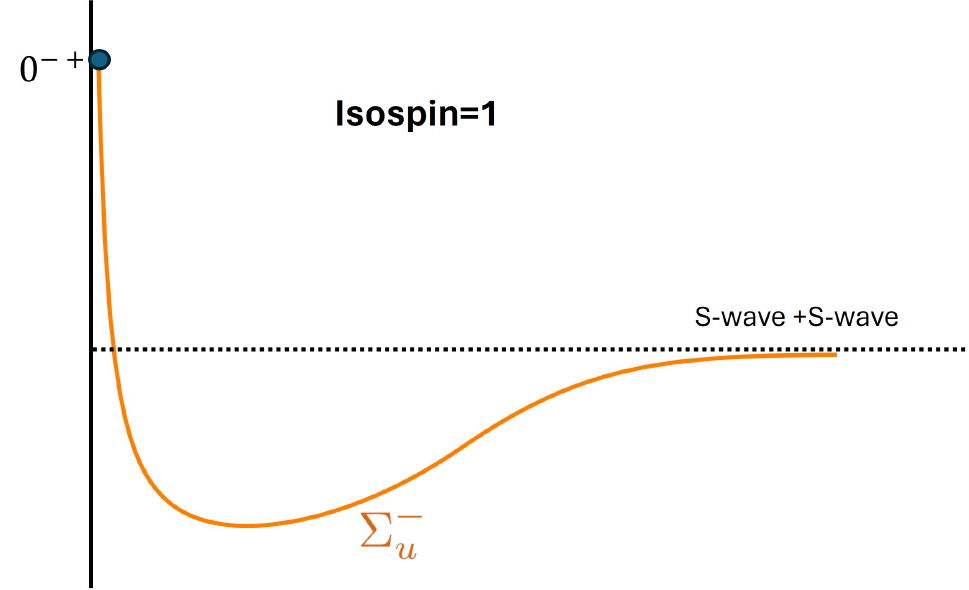}}
\caption{\label{fig3} Possible behaviours towards the open flavor thresholds of the isospin $I=1$ $\Sigma_g^+$, $\Pi_g$ and $\Sigma_u^-$ potentials~\cite{Berwein:2024ztx}.} 
\end{figure}

{\bf Mixing at long distance.}
Due to the string breaking mechanism, at long distance the potentials go into open flavor hadronic states whose light degrees of freedom have the same Born--Oppenheimer quantum numbers, see table~\ref{tab2}.
In dependence on how the Born--Oppenheimer potentials approach the open flavor threshold they may or may not support bound states, which may provide a dynamical explanation of why not all possible tetraquark or pentaquark multiplets
are seen in experiments~\cite{Braaten:2024tbm}.
In figure \ref{fig3}, we show some possible behaviours of the isospin $I=1$ $\Sigma_g^+$, $\Pi_g$ and $\Sigma_u^-$ potentials.
In the first sketch the potential $\Pi_g$ supports a bound state, in the second one it does not.
The potentials $\Sigma_g^+$ and $\Sigma_u^-$ support bound states in all sketches.

\begin{figure}[ht]
\begin{center}
{\epsfxsize=9truecm \epsfbox{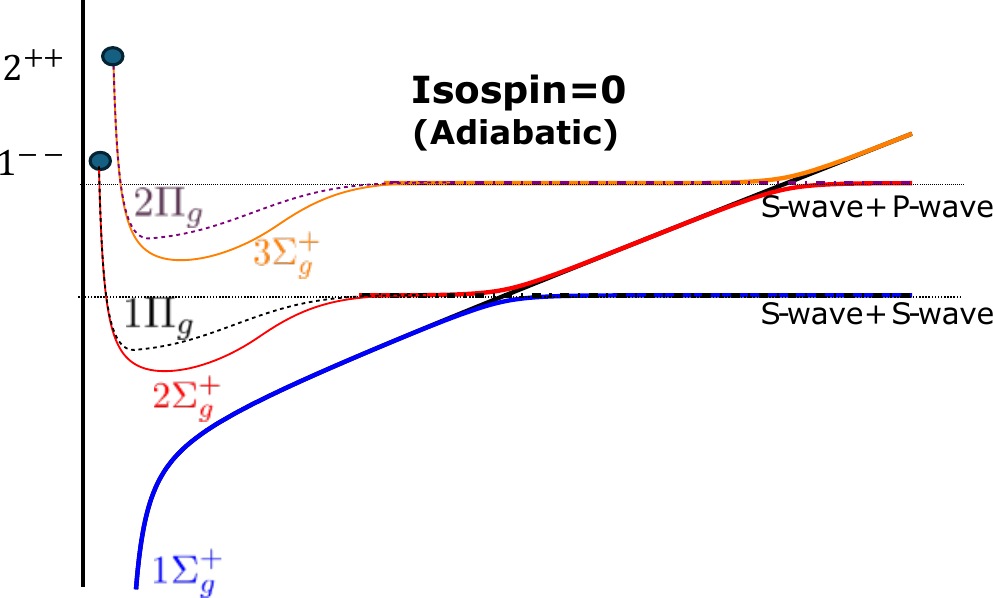}}
\caption{\label{fig4} Avoided level crossing between the quarkonium potential with BO quantum number  $\Sigma_g^+$ and 
the  first two tetraquark potentials with BO quantum numbers  $\Sigma_g^{+\prime}$, $\Sigma_g^{+\prime \prime}$
in the isospin-singlet $I=0$ case~\cite{Berwein:2024ztx}.
The  adiabatic static energies are labeled $1\Sigma_g^+$, $2\Sigma_g^+$, $3\Sigma_g^+$, $1\Pi_g$ and $2\Pi_g$.}
\end{center}
\end{figure}

If mixing between Born--Oppenheimer potentials with the same quantum numbers is accounted for, then the eigenvalues of the potential matrix are called {\it adiabatic potentials}.
These are the physical energies of the static sources at different distances $r$.
In the case of a $2\times 2$ potential matrix with $E_1(r)$ and $E_2(r)$ as diagonal entries and $V_{12}(r)$ as off-diagonal mixing potential, the physical energies $\mathcal{E}_\pm(r)$ read
\begin{equation}
  \mathcal{E}_\pm(r) = \frac{E_1(r) + E_2(r)}{2} \pm \frac{1}{2}\sqrt{(E_1(r)-E_2(r))^2 + 4 V_{12}(r)^2 }.
\label{adiabaticE}  
\end{equation}
If the energy difference $|E_1(r)-E_2(r)|$ is at any distance much larger than $|V_{12}(r)|$, then the effect of the mixing on the spectrum is small and $ \mathcal{E}_{+(-)}(r) \approx  E_{1(2)}(r)$.
If the energy levels $E_1(r)$ and $E_2(r)$ cross at some distance, typically at the string breaking distance of about 1.2~fm, then the phenomenon of {\it avoided level crossing} happens,
where the lower adiabatic energy assumes at large distance the flat threshold behaviour encoded in one of $E_1(r)$ and $E_2(r)$, while the higher adiabatic energy keeps growing.
This is illustrated for the adiabatic isospin $I=0$ potentials in figure~\ref{fig4}.
At short distance, the adiabatic potential $1\Sigma_g^+$ has the attractive behavior of the quarkonium $\Sigma_g^+$ Born--Oppenheimer potential,
while the adiabatic potentials $\left\{2\Sigma_g^+, 1\Pi_g\right\}$ and $\left\{3\Sigma_g^+, 2\Pi_g\right\}$ have the repulsive behavior of the tetraquark Born--Oppenheimer potentials
$\left\{\Sigma_g^{+\prime}, \Pi_g\right\}$ and $\left\{\Sigma_g^{+\prime\prime}, \Pi_g^{\prime}\right\}$ that form degenerate multiplets corresponding to the $1^{--}$ and $2^{++}$ adjoint mesons.
At about 1.2~fm, the $1\Sigma_g^+$ adiabatic potential and the $2\Sigma_g^+$ tetraquark potential are the result of avoided level crossing.
The $2\Sigma_g^+$ potential assumes a linearly rising confining behavior up to the next avoided level crossing.
The $1\Sigma_g^+$ potential goes into the S-wave + S-wave static heavy-light meson-antimeson threshold.
The avoided crossing is not affecting the Born--Oppenheimer potentials $\Pi_g=1\Pi_g$ and the $\Pi_g^\prime=2\Pi_g$.

We observe that, although a small mixing in the absence of avoided level crossing has a small impact on the spectrum, it may be the main responsible for the decay of the state to open flavor hadrons.
This has been exploited in~\cite{Bruschini:2023tmm,TarrusCastella:2024zps,Braaten:2024stn,Brambilla:2025xma}.

Accounting for both the mixing/degeneracy at short distance of the potentials $\Sigma_g^\prime$ and $\Pi_g$ and the mixing/avoided level crossing
of the potentials $\Sigma_g$ and $\Sigma_g^\prime$ at large distance leads to the coupled  Schr\"odinger equations relevant to describe isospin $I=0$ quarkonia and tetraquarks:
\begin{align}
&\left[
-\frac{1}{m_Qr^2}\,\partial_rr^2\partial_r+\frac{1}{m_Qr^2}
{\begin{pmatrix}
l\left(l+1\right) & 0 & 0\\
0                 & l(l+1)+2        & -2\sqrt{l(l+1)} \\
0                 & -2\sqrt{l(l+1)} & l(l+1)
\end{pmatrix}}\right.
\nonumber\\
&\hspace{3.5 cm}\left.
+\begin{pmatrix} V_{\Sigma_{g}^{+}}(r) &  V_{\Sigma_{g}^{+}-\Sigma_{g}^{+\prime}}(r) & 0 \\
    V_{\Sigma_{g}^{+}-\Sigma_{g}^{+\prime}}(r) & V_{\Sigma_{g}^{+\prime}}(r) & 0\\
      0 & 0 & V_{\Pi_g}(r)\end{pmatrix}
      \right]
  \hspace{-4pt}\begin{pmatrix} \psi_{\Sigma_g}^{(N)}    \\ \psi_{\Sigma^{\prime}_g}^{(N)} \\ \psi_{\Pi_g }^{(N)}\end{pmatrix}={\mathcal{E}_N} \begin{pmatrix} \psi_{\Sigma_g}^{(N)} \\ \psi_{\Sigma^{\prime}_g}^{(N)} \\ \psi_{\Pi_g}^{(N)}\end{pmatrix},
\label{coupledI0}
\end{align}
where $V_{\Sigma_{g}^{+}-\Sigma_{g}^{+\prime}}(r)$ is the $\Sigma_{g}^{+}$-$\Sigma_{g}^{+\prime}$ mixing potential.
In particular, for $l=1$ these equations describe $\chi_{c1}(3872)$ and the $\chi_c(1P)$ charmonium state.

\section{Tetraquarks: $\chi_{c1}(3872)$ and $T_{cc}^+(3875)$}
\label{sec:tetra}
{\bf The $\chi_{c1}(3872)$ state.}
The state $\chi_{c1}(3872)$ has been the first observed XYZ state~\cite{Belle:2003nnu,LHCb:2013kgk,LHCb:2015jfc,LHCb:2020fvo}.
Its quantum numbers are $J^{PC} = 1^{++}$ ($I=0$).
The state is very close to the $D^{*0}\bar{D}^0$ threshold: $M_{\chi_{c1}(3872)} - (M_{D^{*0}}+M_{\bar{D}^0}) = -0.07 \pm 0.12$~MeV.
Its most likely quark content is that of a charmonium tetraquark: $c\bar{c}q\bar{q}$, where $q=u$, $d$.
We identify $\chi_{c1}(3872)$ with the $1^{++}$ (isospin $I=0$) state of the lowest $k^{PC}=1^{--}$ multiplet, the one with $l=1$, in table~\ref{tab3}.

\begin{table}[t]
	\begin{tabular}{|c|c|c|c|c|c|}
		\hline
		\multirow{2}{*}{\hspace{2pt}$\begin{array}{c} Q\bar{Q}\\\text{color state}\end{array}$\hspace{2pt}} & \multirow{2}{*}{\hspace{2pt}$\begin{array}{c}\text{Light spin}\\k^{PC}\end{array}$\hspace{2pt}} & \multirow{2}{*}{\hspace{2pt} $\begin{array}{c} \text{BO quantum \#}\\
                  \end{array}$\hspace{2pt}}&\multirow{2}{*}{\hspace{2pt} $l$\hspace{2pt}}& \multirow{2}{*}{\hspace{2pt} $\begin{array}{c}J^{PC}\\\{S_Q=0, S_Q=1\}\end{array}$\hspace{2pt}}& \multirow{2}{*}{\hspace{2pt}Multiplets\hspace{2pt}}\\
		& & & & &  \\
		\hline\hline
	\multirow{6}{*} {\hspace{2pt}$\begin{array}{c} \text{Octet}\\\mathbf{8}\end{array}$\hspace{2pt}}
&\multirow{3}{*}{$0^{-+}$} & \multirow{3}{*}{$\Sigma_u^-$}  & \hspace{2pt}$0$\hspace{2pt} & \hspace{2pt}$\{0^{++}, 1^{+-}\}$\hspace{2pt}
		&\hspace{2pt}$T_1^0$\hspace{2pt}\\
		\cline{4-6}
		& & & \hspace{2pt}$1$\hspace{2pt} & \hspace{2pt}$\{1^{--}, \left(0, 1, 2\right)^{-+}\}$\hspace{2pt}&\hspace{2pt}$T_2^0$\hspace{2pt}\\
		\cline{4-6}
		& & & \hspace{2pt}$2$\hspace{2pt} & \hspace{2pt}$\{2^{++}, \left(1, 2, 3\right)^{+-}\}$\hspace{2pt}&\hspace{2pt}$T_3^0$\hspace{2pt}
		\\
		\cline{2-5}\cline{2-6}
		&\multirow{4}{*}{$1^{--}$} & ${\Sigma_g^{+\prime},\Pi_g}$  & \hspace{2pt}$1$\hspace{2pt} & \hspace{2pt}$\{1^{+-}, (0,1,2)^{++}\}$\hspace{2pt}&\hspace{2pt}$T_1^1$\hspace{2pt} \\
		\cline{3-6}
		& & ${\Sigma_g^{+\prime}}$  & \hspace{2pt}$0$\hspace{2pt} & \hspace{2pt}$\{0^{-+},  1^{--}\}$\hspace{2pt}&\hspace{2pt}$T_2^1$\hspace{2pt} \\
  \cline{3-6}
		& & ${\Pi_g}$  & \hspace{2pt}$1$\hspace{2pt} & \hspace{2pt}$\{1^{-+},  (0,1,2)^{--}\}$\hspace{2pt}&\hspace{2pt}$T_3^1$\hspace{2pt} \\
		\cline{3-6}
		& & ${\Sigma_g^{+\prime},\Pi_g}$  & \hspace{2pt}$2$\hspace{2pt} & \hspace{2pt}$\{2^{-+}, (1,2, 3)^{--}\}$\hspace{2pt} &\hspace{2pt}$T_4^1$\hspace{2pt}\\
		\hline
	\end{tabular}
        \caption{\label{tab3} $J^{PC}$ multiplets for the lowest $Q\bar{Q}q\bar{q}$ tetraquarks.
          $S_Q$ is the spin of the heavy quark pair.
          The third column shows the BO quantum numbers of the  potentials appearing in the Schr\"odinger equations of the respective multiplet.}
\end{table}

According to table~\ref{tab3}, $\chi_{c1}(3872)$ is the solution of the coupled Schr\"odinger equations \eqref{coupledI0} for $\Sigma_g^+$, $\Sigma_g^{+\prime}$ and $\Pi_g$.
The adiabatic potentials $V_{1\Sigma_{g}^{+}}(r)$, $V_{2\Sigma_{g}^{+}}(r)$ and $V_{1\Pi_{g}}(r)$, sketched in figure~\ref{fig4},
and the mixing potential $V_{\Sigma_{g}^{+}-\Sigma_{g}^{+\prime}}(r)$ may be extracted from lattice QCD data~\cite{Bulava:2019iut,Bulava:2024jpj}
supplemented by symmetry constraints at short and long distances, where lattice data are not available.

\begin{figure}
\begin{center}    
{\epsfxsize=9truecm \epsffile{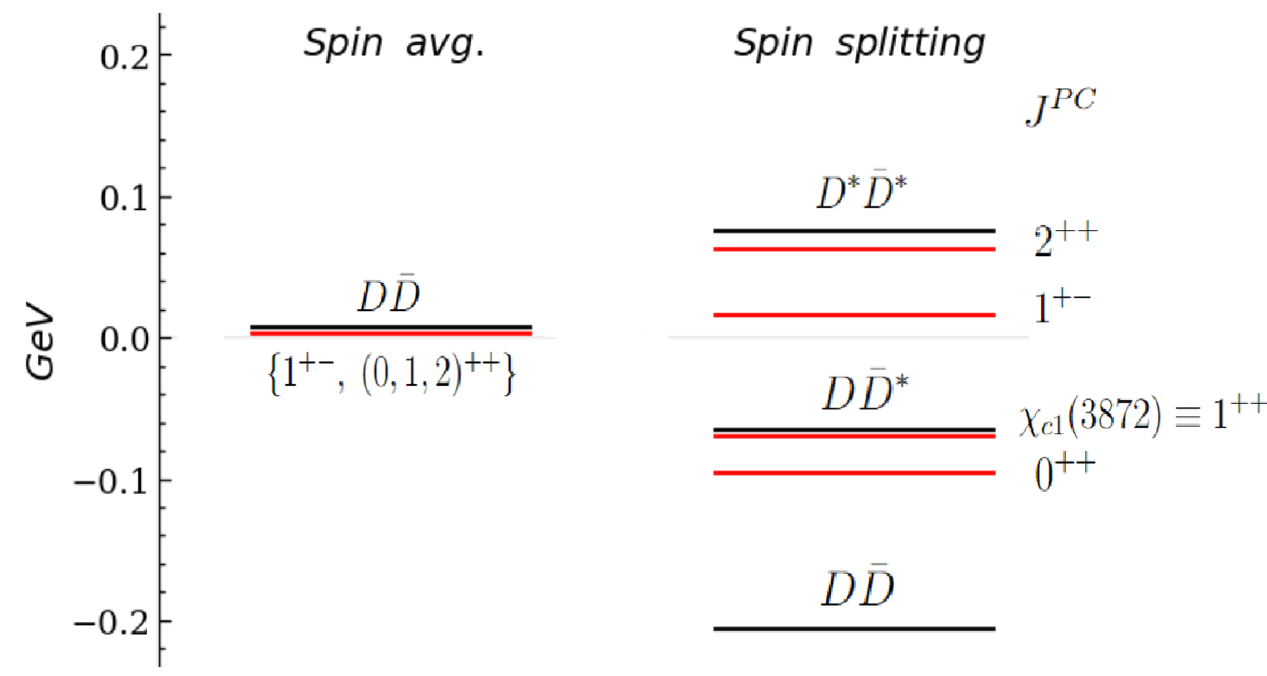}}
\caption{\label{fig5} Mass spectrum of states in the spin multiplet of $\chi_{c1}(3872)$ (red lines) vs thresholds (black lines)~\cite{Brambilla:2024thx}. }
\end{center}
\end{figure}

After using all constraints, the potentials depend on only one free parameter: the adjoint meson mass $\Lambda^a_{1^{--}}$.
For the critical value $\Lambda^a_{1^{--}} \approx 914$ MeV, one finds a $1^{++}$ state with mass 3872 MeV that can be identified with the $\chi_{c1}(3872)$~\cite{Brambilla:2024thx}.
In the absence of more direct information, spin splittings are taken from lattice hybrid spin splittings.
Other members of the multiplet are a $1^{+-}$ state with mass $3957(11)$~MeV, a $0^{++}$ state with mass $3846(11)$~MeV, and a $2^{++}$ state with mass $4004(14)$~MeV, see figure~\ref{fig5}.
The $1^{+-}$ state could be a candidate for $X(3940)$ seen by Belle (not confirmed).
In the spin-averaged case we find a deeper bound state with mass $3537$~MeV that we identify with the spin-averaged $\chi_{c}\left(1P\right)$ state.

We find that the scattering length of $\chi_{c1}(3872)$ is about 15~fm and that its quarkonium component, $|\psi_{\Sigma_g^+}|^2$, is about $8\%$, 
its $\Sigma_g^{+\prime}$ tetraquark component, $|\psi_{\Sigma_g^{+\prime}}|^2$, is about $38\%$,
and its $\Pi_g$ tetraquark component, $|\psi_{\Pi_g}|^2$, is about $54\%$.
From this, it follows (assuming radiative decays via quarkonium component)
\begin{equation}
\frac{\Gamma_{\chi_{c1}(3872) \to \gamma \psi(2S)}}{\Gamma_{\chi_{c1}(3872) \to \gamma J/\psi}} = 2.99 \pm 2.36,
\end{equation}
which is consistent with the LHCb measure $1.67 \pm 0.25$ \cite{LHCb:2024tpv}.

A similar analysis can be done in the bottomonium sector.
There, using the same potential tuned in the charmonium sector, we find that the lowest spin multiplet states $1^{++}$, $1^{+-}$, $0^{++}$ and $2^{++}$ get masses
about 10595 MeV, 10612 MeV, 10586 MeV and 10627 MeV, respectively~\cite{Brambilla:2024thx}.
Note that the $\chi_{b1}$ $1^{++}$ state that we find, differently from the corresponding $\chi_{c1}(3872)$ state in the charmonium sector, is not on a meson-antimeson threshold.

{\bf The $T_{cc}^+(3875)$ state.}
The state $T_{cc}^+(3875)$ has been the first observed doubly charmed tetraquark at LHCb~\cite{LHCb:2021vvq,LHCb:2021auc}.
Its quantum numbers are $J^{P} = 1^{+}$ ($I=0$).
This state is so far the longest living exotic particle: $\Gamma \approx 50$ keV.
It lies below the $D^{*+}D^0$ threshold: $M_{T_{cc}^+(3875)} - (M_{D^{*+}}+M_{D^0}) = -0.27 \pm 0.06$~MeV.
Its most likely quark content is that of a charmonium tetraquark: $cc\bar{u}\bar{d}$.
We identify $T_{cc}^+(3875)$ with the $1^{+}$ ($I=0$) state of the lowest $k^{P}=0^{+}$ multiplet, the one with $l=0$ and $S_Q=1$, in table~\ref{tab4}.
The $1^+$ state of $\Sigma_g^+$ is the lowest state if $V_{\Sigma_{g}^{+}}(r)$ lies below $V_{\Sigma_{g}^{-}}(r)$ and $V_{\Pi_{g}}(r)$.
We consider only $QQ$ pairs in a color antitriplet configuration because the sextet is repulsive and therefore presumably associated with higher states.

\begin{table}[t]
\begin{center}
  \begin{tabular}{|c|c|c|c|c|c|c|}
		\hline
		\multirow{2}{*}{\hspace{2pt}$\begin{array}{c} QQ\\\text{ color state}\end{array}$\hspace{2pt}} & \multirow{2}{*}{\hspace{2pt}$\begin{array}{c} \text{Light spin}\\ k^{P}\end{array}$\hspace{2pt}} & \multirow{2}{*}{\hspace{2pt} $\begin{array}{c} \text{BO quantum \#}\\
                  \end{array}$\hspace{2pt}} & \multirow{2}{*}{\hspace{2pt}$\begin{array}{c} \text{Isospin}\\I\end{array}$ \hspace{2pt}}
                & \multirow{2}{*}{\hspace{2pt} $l$\hspace{2pt}} & \multicolumn{2}{|c|}{$J^{P}$}\\
		\cline{6-7}
		& & & & & \hspace{2pt}$S_Q=0$\hspace{2pt} & \hspace{2pt}$S_Q=1$\hspace{2pt} \\
		\hline\hline
		\multirow{4}{*}{\hspace{2pt}$\begin{array}{c} \text{Antitriplet}\\\bar{\mathbf{3}}\end{array}$\hspace{2pt}} &\multirow{2}{*}{$0^+$} & \multirow{2}{*}{${\Sigma_g^+}$} & \multirow{2}{*}{$0$}& $0$ & \hspace{2pt} --- & \hspace{2pt}$1^+$\hspace{2pt} \\
		\cline{5-7}
		& & & & 1 &\hspace{2pt}$1^-$\hspace{2pt}&---\\
		\cline{2-7}
		& \multirow{2}{*}{$1^+$} & \multirow{2}{*}{${\Sigma_g^-, \Pi_g}$} & \multirow{2}{*}{$1$}& $0$ &  \hspace{2pt}$0^-$\hspace{2pt}& --- \\
		\cline{5-7}
		& & & & 1 & $1^{-}$&\hspace{2pt}$\left(0, 1, 2\right)^+$\hspace{2pt}\\
		\cline{5-7}
		\hline
         \end{tabular}
	\caption{\label{tab4} $J^P$ multiplets for the lowest $QQ\bar{q}\bar{q}$ tetraquarks.} 
\end{center}
\end{table}

According to table~\ref{tab4}, $T_{cc}^+(3875)$ is the solution of a single radial Schr\"odinger equation with the Born--Oppenheimer potential $\Sigma_g^+$
that approaches a $0^+$ triplet meson at short and a meson-meson $I=0$ pair at large distance~\cite{Berwein:2024ztx}:
\begin{equation}
\left[-\frac{1}{m_Qr^2}\,\partial_r\,r^2\,\partial_r+\frac{l(l+1)}{m_Qr^2}+V_{\Sigma_{g}^+}(r)\right]\psi_{\Sigma_g^+}^{(N)}=\mathcal{E}_N\,\psi_{\Sigma_g^+}^{(N)}.
\end{equation}
The potential $V_{\Sigma_{g}^{+}}(r)$ can be extracted from lattice QCD ~\cite{Lyu:2023xro,Bicudo:2024vxq}.

In the analysis of~\cite{Brambilla:2024thx}, we get a $T_{cc}$ state 323 keV below the $DD^*$ threshold that we identify with $T_{cc}^+(3875)$ by fixing the $0^+$ triplet meson mass $\Lambda^t_{0^+}$ to be 664 MeV.
The scattering length turns out to be about 8 fm.

Also in this case, a similar analysis can be done in the bottomonium and in the $bc$ sectors.
Results are consistent with independent calculations in lattice NRQCD and QCD~\cite{Brambilla:2024thx}.

\section{Pentaquarks}
Four states with isospin $1/2$ and unknown $J^P$, $P_{c\bar{c}}(4312)^+$, $P_{c\bar{c}}(4380)^+$, $P_{c\bar{c}}(4440)^+$ and $P_{c\bar{c}}(4457)^+$,
have been observed at LHCb~\cite{LHCb:2019kea}.
Also 2 states with isospin 0 were found, $P_{c\bar{c}s}(4338)^0$, $P_{c\bar{c}s}(4459)^0$, but we limit our present discussion to the isospin $1/2$ states.

There are two thresholds relevant for the lowest lying pentaquarks.
These are the $\Lambda_c\bar{D}$ ($\Lambda_b\bar{B}$ in the bottomonium case) threshold, where the light quark pair in the baryon has spin 0,
and the $\Sigma_c\bar{D}$ ($\Sigma_b\bar{B}$) threshold, where the light quark pair in the baryon has spin 1.
The threshold $\Lambda_c\bar{D}$ ($\Lambda_b\bar{B}$) may be the long range tail of a $\left(1/2\right)_g$ Born--Oppenheimer potential joining at short range with an adjoint baryon $(1/2)^+$ mass.
The threshold $\Sigma_c\bar{D}$ ($\Sigma_b\bar{B}$) may be the long range tail of a $\left(1/2\right)_g$ Born--Oppenheimer potential joining at short range with an adjoint baryon $(1/2)^+$ mass 
or the long range tail of the two $\left\{\left(1/2\right)^\prime_g,\left(3/2\right)_g \right\}$ Born--Oppenheimer potentials becoming degenerate at short range with an adjoint baryon $(3/2)^+$ mass.
In order to describe the lowest lying pentaquarks in the BOEFT framework, we need information about the above four Born--Oppenheimer potentials.
None of these potentials has been computed in lattice QCD.
Hence, all conclusions that we can draw at this point are based on assumptions.

\begin{table}[ht]
\begin{center}
  \begin{tabular}{|c|c|c|c|c|}
		\hline
                \multirow{2}{*}{\hspace{2pt}$\begin{array}{c} Q\bar{Q}\\\text{color state}\end{array}$\hspace{2pt}}
                & \multirow{2}{*}{\hspace{2pt}$\begin{array}{c}\text{Light spin}\\k^{P}\end{array}$\hspace{2pt}}
                & \multirow{2}{*}{\hspace{2pt} $\begin{array}{c}\text{BO quantum \#}\\
                  \end{array}$\hspace{2pt}} &\multirow{2}{*}{\hspace{2pt} $l$\hspace{2pt}}
                & \multirow{2}{*}{\hspace{2pt} $\begin{array}{c}J^{P}\\\{S_Q=0, S_Q=1\}\end{array}$\hspace{2pt}}\\
		& &  &  &  \\
		\hline\hline
		\multirow{2}{*} {\hspace{2pt}$\begin{array}{c} \text{Octet}\\\mathbf{8}\end{array}$\hspace{2pt}}
                &\multirow{1}{*}{$(1/2)^{+}$}
                &\multirow{1}{*}{$\left(1/2\right)_g$}\hspace{2pt}
                &\hspace{2pt}$1/2$\hspace{2pt}
                & \hspace{2pt}$\{1/2^{-}, \left(1/2, 3/2\right)^-\}$\hspace{2pt}
		\\
		\cline{2-5}
		&\multirow{1}{*}{$(3/2)^{+}$}
                &\multirow{1}{*}{$\left(1/2\right)_g^{\prime}$, $\left(3/2\right)_g$}\hspace{2pt}
                &\hspace{2pt}$3/2$\hspace{2pt} & \hspace{2pt}$\{3/2^{-}, (1/2, 3/2, 5/2)^{-}\}$\hspace{2pt}\\
		\hline
	\end{tabular}
\caption{\label{tab5} $J^{P}$ multiplets for the lowest pentaquark states $Q\bar{Q}qqq$. 
  The $Q\bar{Q}$ pair is in a color octet configuration.
  The BO quantum numbers are written in the third column as $\left(\Lambda\right)_\eta$, where $\eta = g$ denotes positive parity.}
\end{center}
\end{table}

If all four potentials would support bound states, i.e. connecting to the thresholds from below (at short distance all the potentials behave as repulsive octet potentials),
then there would be ten low-lying pentaquark states.
Ten low-lying pentaquark states are predicted in compact pentaquark models such as in \cite{Ali:2019npk}.

The fact that there is no experimental evidence for states near the $\Lambda_c\bar{D}$ threshold
could be explained by the Born--Oppenheimer potential connecting to that threshold falling off monotonically from above and therefore not supporting bound states.
In such a scenario, bound states are supported only by the three Born--Oppenheimer potentials connecting to the $\Sigma_c\bar{D}$ (and in the bottomonium sector $\Sigma_b\bar{B}$) threshold.
These are the seven states grouped in the spin multiplets listed in table~\ref{tab5}.
This scenario has been analyzed in~\cite{Brambilla:2025xma}. 
Clearly, in such a scenario three of the seven expected low-lying pentaquarks still need to be detected.

\begin{figure}[ht]
\begin{center}    
{\epsfxsize=9truecm \epsffile{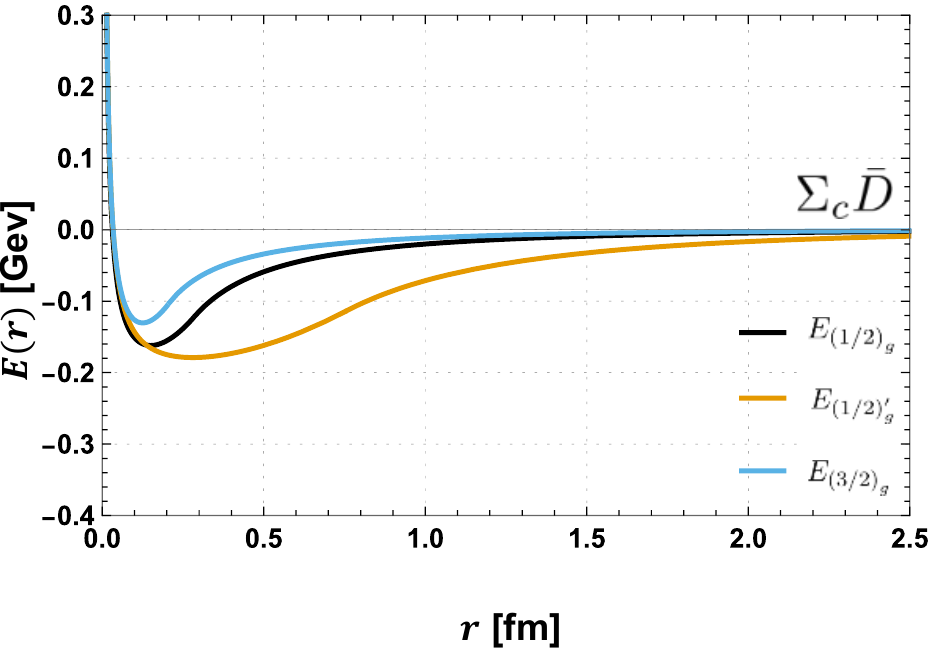}}
\caption{\label{fig6} Born--Oppenheimer potentials for the lowest-lying pentaquarks in the scenario of~\cite{Brambilla:2025xma}.}
\end{center}
\end{figure}

\begin{figure}[ht]
\begin{center}    
{\epsfxsize=8truecm \epsffile{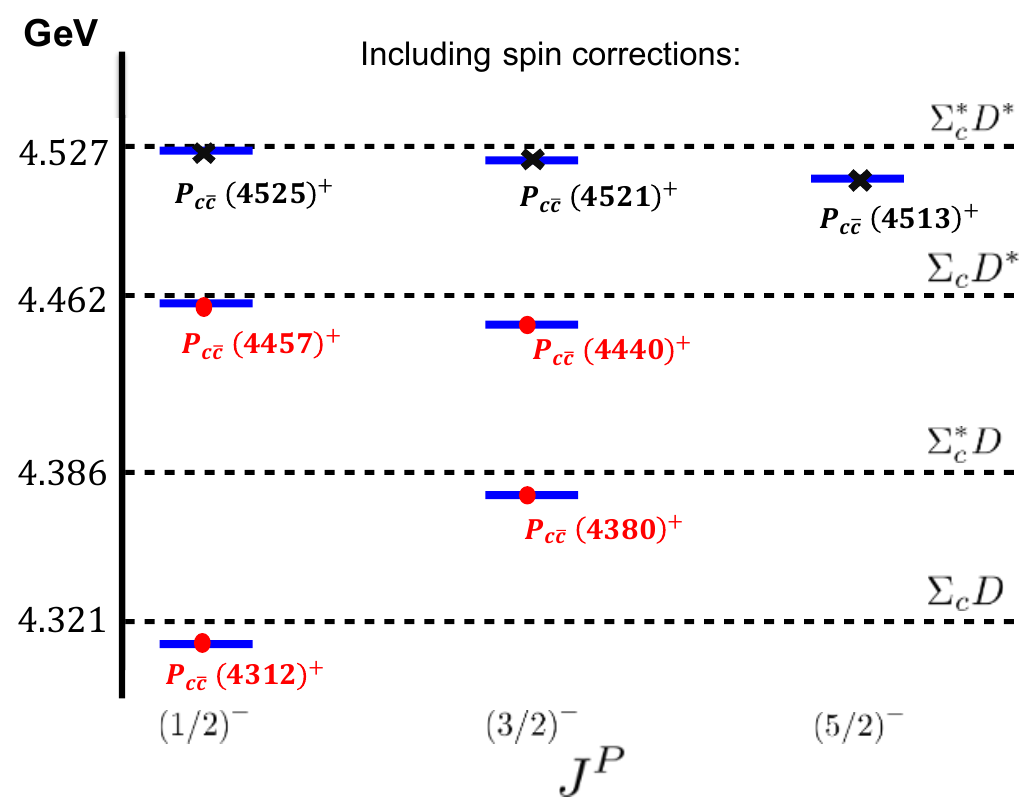}}
\caption{\label{fig7} Lowest lying charmonium pentaquark multiplets supported by the potentials in figure~\ref{fig6}, from~\cite{Brambilla:2025xma}.
Experimentally detected states are marked in red.}
\end{center}
\end{figure}

\begin{figure}[ht]
\begin{center}    
{\epsfxsize=8truecm \epsffile{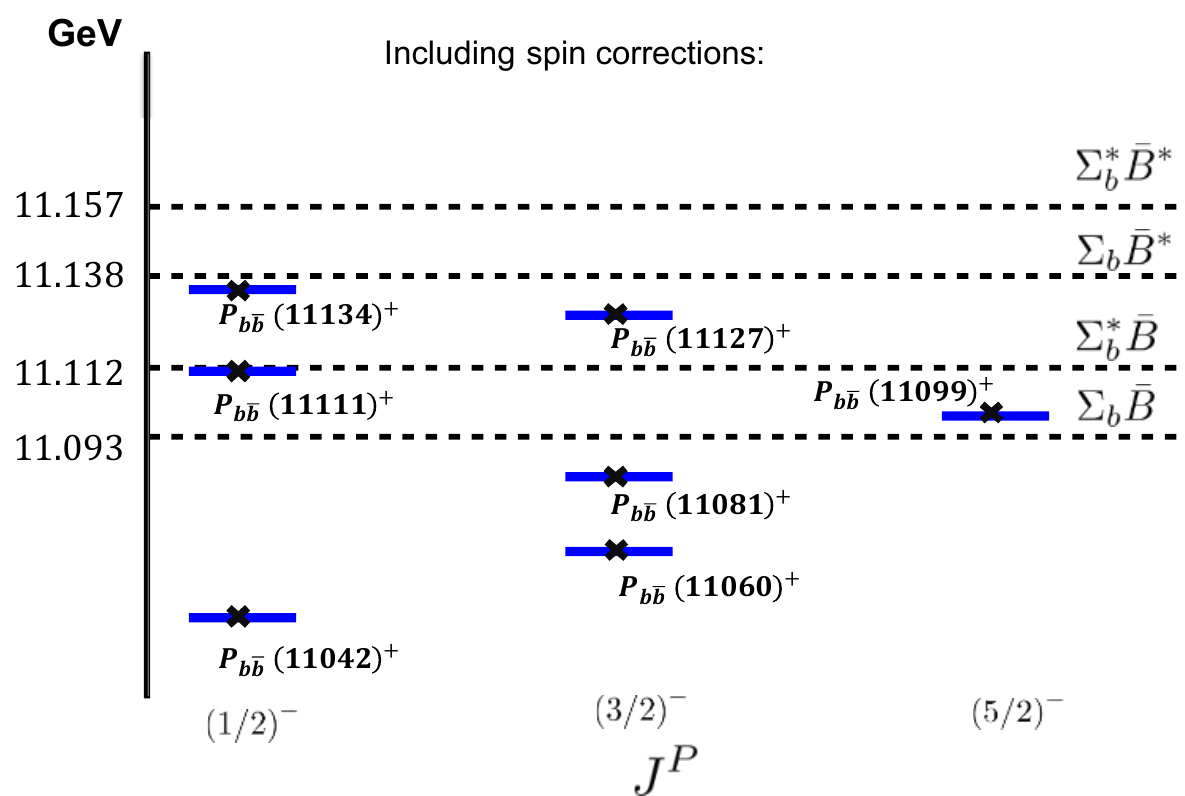}}
\caption{\label{fig8} Lowest lying bottomonium pentaquark multiplets supported by the potentials in figure~\ref{fig6} approaching now the  $\Sigma_b\bar{B}$ threshold,
  from~\cite{Brambilla:2025xma}.}
\end{center}
\end{figure}

According to table~\ref{tab5}, pentaquarks with $k^P = (1/2)^+$ are solutions of a single radial Schr\"odinger equation
involving the Born--Oppenheimer potential $(1/2)_g$:
\begin{align}
  \left[-\frac{1}{m_Qr^2}\,\partial_r\,r^2\,\partial_r+\frac{(l-1/2)(l+1/2)}{m_Qr^2}+V_{(1/2)_g}(r)\right]\psi_{(1/2)^+}^{(N)}=
  \mathcal{E}_N\,\psi_{(1/2)^+}^{(N)},
\end{align}
whereas pentaquarks with $k^P = (3/2)^+$ are solutions of coupled radial Schr\"odinger equations
involving the Born--Oppenheimer potentials $(1/2)_g^\prime$ and $(3/2)_g$: 
\begin{align}
  &\Bigg[-\frac{1}{m_Qr^2}\,\partial_rr^2\partial_r+\frac{1}{m_Qr^2}
    \begin{pmatrix} l(l-1)+9/4 & -\sqrt{3l(l+1) - 9/4} \\ -\sqrt{3l(l+1) -9/4} & l(l+1) -3/4\end{pmatrix} \nonumber\\
      &\hspace{2.5 cm}+\begin{pmatrix} V_{(1/2)^\prime_g}(r) & 0 \\ 0 & V_{(3/2)_g}(r) \end{pmatrix}\Bigg]\hspace{-4pt}
  \begin{pmatrix} \psi_{(1/2)^{+\prime}}^{(N)} \\ \psi_{(3/2)^+}^{(N)}\end{pmatrix}=
  \mathcal{E}_N\begin{pmatrix} \psi_{(1/2)^{+\prime}}^{(N)} \\ \psi_{(3/2)^+}^{(N)}\end{pmatrix}.
\end{align}
Parameterizing the three Born--Oppenheimer potentials to overlap with the threshold from below at large distance and with the octet potential at short distance reduces their uncertainties
to just two unknown parameters, the adjoint baryon masses $\Lambda_{(1/2)^+}^b$ and $\Lambda_{(3/2)^+}^b$.
There are two possible choices of the adjoint baryon masses able to reproduce the four experimentally proved pentaquark states.
The preferable choice assigns $\Lambda_{(1/2)^+}^b \approx 1125$ MeV and $\Lambda_{(3/2)^+}^b \approx 1152$ MeV (in the RS scheme~\cite{Bali:2003jq}).
This choice is preferable because it leads, in particular, to a decay width for $P_{c\bar{c}}(4312)^+ \to J/\psi + X$ that is consistent with the total width of the state measured by LHCb; 
the other choice does not.
The relevant Born--Oppenheimer potentials in this scenario are shown in figure~\ref{fig6}.
Note that all the potentials approach (from below) the $\Sigma_c\bar{D}$ threshold at large distances
and that the $\left\{\left(1/2\right)^\prime_g,\left(3/2\right)_g \right\}$ potentials are degenerate at short distance.
The resulting charmonium pentaquark spectrum is shown in figure~\ref{fig7}.
To the pentaquark states are assigned specific $J^P$ quantum numbers, in particular $J^P=(3/2)^-$ to $P_{c\bar{c}}(4440)^+$ and $J^P=(1/2)^-$ to $P_{c\bar{c}}(4457)^+$.
Seven low lying pentaquark states are also predicted by molecular models~\cite{Liu:2019tjn,Du:2019pij,Xiao:2019aya,Du:2021fmf}.
Using the same Born--Oppenheimer potentials tuned on the charmonium pentaquark spectrum, see figure~\ref{fig6}, for the lowest lying multiplets in the bottomonium pentaquark spectrum 
close to the $\Sigma_b\bar{B}$ threshold, we can predict the lowest lying bottomonium pentaquark states.
These are shown in figure~\ref{fig8}.

A final scenario consists in assuming that only the potentials with BO quantum numbers $\{\left(1/2\right)_g^\prime, \left(3/2\right)_g\}$
cross the $\Sigma_c\bar{D}$ ($\Sigma_b\bar{B}$) threshold and then approach it from below, supporting bound states, while the potential with BO quantum number
$\left(1/2\right)_g$ joins the $\Sigma_c\bar{D}$ ($\Sigma_b\bar{B}$) threshold from above, like the $\left(1/2\right)_g$ Born--Oppenheimer potential 
is assumed to do with the $\Lambda_c\bar{D}$ ($\Lambda_b\bar{B}$) threshold. 
This scenario implies that the lowest pentaquark states are only four in agreement with current observations.
This case has been studied in \cite{Alasiri:2025roh}.

\section{Quarkonium hybrids}
The description of quarkonium hybrids in the framework of BOEFT is much more constrained than the description of tetraquarks and pentaquarks.
The reason is that  there is an extended literature dealing with the computation of quarkonium hybrid Born--Oppenheimer potentials on the lattice,
mostly in the SU(3) gauge theory~\cite{Juge:2002br,Bali:2003jq,Capitani:2018rox}, but recently also with $3+1$ dynamical fermions~\cite{Hollwieser:2025nvv}.
This has allowed to develop BOEFT for quarkonium hybrids much earlier than for tetraquarks, doubly heavy baryons or pentaquarks~\cite{Berwein:2015vca}.

The lowest lying quarkonium hybrid Born--Oppenheimer potentials have BO quantum numbers $\Pi_u$ and $\Sigma_u^-$. 
They become degenerate with the $k^{PC} = 1^{+-}$ gluelump mass in the short distance limit.
The lowest lying quarkonium hybrid multiplets associated with the $\Sigma_u^-$ and $\Pi_u$ Born--Oppenheimer potentials are listed in table~\ref{tab6}.

\begin{table}[ht]
\begin{center}
  \begin{tabular}{|c|c|c|c|}
 \hline
  \text{BO quantum \#}  & $\,l\,$ & $J^{PC}\{S_Q=0,S_Q=1\}$  & Multiplets          \\  \hline\hline
  $\Sigma_u^-$, $\Pi_u$  & $1$     & $\{1^{--},(0,1,2)^{-+}\}$ & $H_1$  \\
  $\Pi_u$               & $1$     & $\{1^{++},(0,1,2)^{+-}\}$  & $H_2$           \\
  $\Sigma_u^-$  & $0$     & $\{0^{++},1^{+-}\}$         & $H_3$       \\
  $\Sigma_u^-$, $\Pi_u$   & $2$     & $\{2^{++},(1,2,3)^{+-}\}$  & $H_4$ \\
  $\Pi_u$   & $2$     & $\{2^{--},(1,2,3)^{-+}\}$         & $H_5$    \\
  \hline
 \end{tabular}
 \caption{\label{tab6} Lowest $J^{PC}$ quarkonium hybrid multiplets associated with the $\Sigma_u^-$ and $\Pi_u$ Born--Oppenheimer potentials.}
\end{center}
\end{table}

Based on the $\Pi_u$ and $\Sigma_u^-$ Born--Oppenheimer potentials computed on the lattice, the mass spectrum of the lowest lying charmonium and bottomonium hybrid states
listed in table~\ref{tab6} has been computed by solving the coupled Schr\"odinger equations \eqref{Sch_short1} and \eqref{Sch_short2} in~\cite{Berwein:2015vca,Brambilla:2022hhi}.
Since lattice SU(3) gauge theory results have been used, the potentials and the ensuing Schr\"odinger equations do not account for mixings with the open flavor thresholds. 
The fine and hyperfine structure of the charmonium and bottomonium hybrid spectrum has been computed in~\cite{Brambilla:2018pyn,Brambilla:2019jfi,Soto:2023lbh} as it is possible
to infer the behaviour of the spin dependent potentials at short distance from the multipole expansion and at large distance from the effective string theory. 
For quarkonium hybrids the spin dependent potentials are only $1/m_Q$ suppressed~\cite{Oncala:2017hop}.
The description of the fine and hyperfine structure one gets in this way matches remarkably well direct lattice QCD determinations~\cite{Cheung:2016bym,Ryan:2020iog}.

Nevertheless, computing the spectrum is not enough to identify unambiguously quarkonium hybrids among the XYZ states.
Strong constraints are provided by quarkonium hybrid to quarkonium transitions, whose widths have to be smaller than the measured widths of the states.
A similar constraint has been used to favor one solution of the Schr\"odinger equations with respect to another one in the pentaquark case.
Quarkonium hybrid to quarkonium transitions have been computed in~\cite{Oncala:2017hop,TarrusCastella:2021pld,Brambilla:2022hhi}
and the implications for the identification of some of the XYZ states with states possessing a large quarkonium hybrid component have been discussed there in detail.

As we have done for tetraquarks and pentaquarks, the effect of open flavor thresholds can be systematically embedded in the description of hybrids by accounting for
the mixing with suitable Born--Oppenheimer tetraquark potentials.
Since an $S$-wave + $S$-wave meson-antimeson pair can have $k^{PC}=0^{-+}$,
in the Born--Oppenheimer framework this threshold is embedded in the tetraquark potential $V_{\Sigma_u^-}(r)$ (here $V_{\Sigma_u^{-\prime}}(r)$ to distinguish it
from the hybrid potential), see table~\ref{tab2}.
The coupled Schr\"odinger equations describing hybrids, $\psi_{\Pi_u}^{(N)}$ and $\psi_{\Sigma_u}^{(N)}$, mixing with tetraquarks, $\psi_{\Sigma^\prime_u}^{(N)}$, read 
\begin{align}
&\left[-\frac{1}{m_Qr^2}\,\partial_rr^2\partial_r+\frac{1}{m_Qr^2}\begin{pmatrix} l(l+1)+2 & -2\sqrt{l(l+1)} & 0 \\ -2\sqrt{l(l+1)} & l(l+1) & 0 \\ 0 & 0 & l\left(l+1\right) \end{pmatrix}\right.\nonumber\\
&\hspace{1.5 cm}\left.+\begin{pmatrix} V_{\Sigma_{u}^{-}}(r) & 0 & V_{\Sigma_{u}^{-}-\Sigma_{u}^{-\prime}}(r) \\
      0 & V_{\Pi_{u}}(r) & 0\\
      V_{\Sigma_{u}^{-}-\Sigma_{u}^{-\prime}}(r) & 0 & V_{\Sigma_u^{-\prime}}(r)\end{pmatrix}\right]\hspace{-4pt}\begin{pmatrix} \psi_{\Sigma_u}^{(N)} \\ \psi_{\Pi_u}^{(N)} \\ \psi_{\Sigma^\prime_u}^{(N)}\end{pmatrix}={\cal E}_N\begin{pmatrix} \psi_{\Sigma_u}^{(N)} \\ \psi_{\Pi_u}^{(N)}\\ \psi_{\Sigma^\prime_u}^{(N)}\end{pmatrix}.
\end{align}
The potentials $V_{\Sigma_{u}^{-}}(r)$ and $V_{\Pi_{u}}(r)$ are the lowest hybrid potentials, while the mixing potential between
the hybrid $\Sigma_u^-$ and the tetraquark  $\Sigma_u^{-\prime}$, $V_{\Sigma_{u}^{-}-\Sigma_{u}^{-\prime}}(r)$, is so far unknown, i.e. not computed in lattice QCD.
Because hybrids of the multiplets $H_1$, $H_3$, $H_4$, ... are in part or completely excitations of the $\Sigma_u^-$ potential, see table \ref{tab6}, 
they couple to the tetraquark $\Sigma_u^{-\prime}$ and therefore are allowed to decay into $S$-wave meson-meson pairs.
The decay of quarkonium hybrids into two $S$-wave heavy-light mesons is, therefore, unsuppressed~\cite{Bruschini:2023tmm,TarrusCastella:2024zps,Braaten:2024stn}.

\section{Threshold effects in quarkonium}
Thresholds effects in quarkonium are embedded in tetraquark potentials that have the same BO quantum numbers as quarkonium, i.e. $\Sigma_g^+$.
Hence, threshold effects in the quarkonium wavefunction and spectrum are accounted for by solving with respect to
the quarkonium state the coupled Schr\"odinger equations involving 
the quarkonium Born--Oppenheimer potential with quantum numbers $\Sigma_g^+$ and the tetraquark ones with quantum numbers $\Sigma_g^{+\prime}$ and $\Pi_g$,
which overlap at short distance with the $1^{--}$ adjoint meson mass and at large distance with the isospin $I=0$ $k^{PC}=1^{--}$ S-wave plus S-wave meson-antimeson threshold.
These coupled Schr\"odinger equations are those in~\eqref{coupledI0}.

\begin{table}
\begin{center}
\begin{tabular}{|c|c|c|c|c|}  \hline
$n l \, (b \bar{b})$ & $\%  \; \Sigma_g^+$ & $\% \; \Sigma_g^{+'}$  & $\% \; \Pi_g$ & $\Delta E_{nl}\,(\mathrm{MeV})$ \\
\hline\hline
$1S$         & $100$  & $\approx 0$ &   &  $-0.1$ \\
$2S$         & $99.9$  & $0.01$  &   & $-1.0$\\
$3S$         & $99.1$  & $0.9$   &   & $-3.5$\\
$4S$         & $78.8$  & $21.2$ &    & $-16.0$\\ \hline
$1P$         & $99.9$  & $0.1$  & $\approx 0$  & $-0.5$\\
$2P$         & $99.5$  & $0.5$  & $\approx 0$  & $-2.3$\\
$3P$         & $95.8$  & $4.1$  & $0.1$ & $-7.6$\\
$X_b$        & $1.5$  & $44.9$  & $53.6$ &  \\\hline
$1D$         & $99.8$  & $0.2$  & $\approx 0$  & $-1.2$\\
$2D$         & $98.7$  & $1.3$  & $\approx 0$  & $-4.2$\\ \hline
\end{tabular}
\caption{\label{tab7} Threshold effects on the bottomonium spin averaged states from~\cite{Brambilla:2025xxx}.
  $\Delta E_{nl}$ is the mass shift due to threshold effects,
  $\%  \; \Sigma_g^+$,  $\% \; \Sigma_g^{+'}$  and $\% \; \Pi_g$ provide the fractions of state that are in a $\Sigma_g^+$,  $\Sigma_g^{+'}$  and $\Pi_g$ configuration, respectively.}
\end{center}
\end{table}

The results obtained in~\cite{Brambilla:2025xxx} from solving eq.~\eqref{coupledI0} with respect to the quarkonium state,
after having fixed the potentials and the mixing potential as discussed in section~\ref{sec:tetra},
are shown for the bottomonium case in table~\ref{tab7}.
The tetraquark component turns out to be negligible for states well below threshold, but amounts at about $20\%$ for the $\Upsilon(4S)$.
It is clearly the dominant component for the state that we have dubbed $X_b$,
which is the equivalent in the bottomonium spectrum of the spin averaged $\chi_{c1}(3872)$ multiplet in the charmonium spectrum.
In the Born--Oppenheimer framework, but with somewhat different settings, similar studies have been performed in~\cite{Bicudo:2019ymo,TarrusCastella:2022rxb,Bruschini:2025paj}.

\section{Conclusions}
The Born--Oppenheimer EFT is an EFT of QCD able to describe quarkonia, hybrids, doubly heavy baryons, tetraquarks and pentaquarks in a complete and consistent manner
at the {\it soft scale} of bound-state formation.
With respect to the scale hierarchy introduced at the beginning, the soft scale corresponds to the scale $m_Qv$.

Most of the present limitations come from the rather incomplete knowledge of the potentials entering the coupled Schr\"odinger equations.
These should be provided by lattice QCD.
The situation is best for hybrids, although relevant tetraquark potentials are currently being computed by several collaborations. 
The knowledge of the potentials is fundamental.
It may help to discriminate between different scenarios and explain on a dynamical basis why some bound states exist and some possible others do not.
We have seen explicitly how this works in the case of the lowest lying pentaquarks.
If only the lowest lying Born--Oppenheimer potentials are able to support bound states,
this may explain why we see less tetraquark and pentaquark states than we would naively expect just by combining four or five quark states near the different thresholds.

It is worth remarking that in the Born--Oppenheimer framework one does not need to make a priori assumptions on the dominant configuration that the quarks are going to assume in the bound state
(compact tetraquark, molecule, hadroquarkonium, ...).
It is ultimately the solution of the coupled Schr\"odinger equations that fixes the content of the states, their binding energies and radii.

The Born--Oppenheimer EFT still needs to be completed at the {\it ultrasoft scale}, which is the scale of order $m_Qv^2$ or smaller where low energy degrees of freedom live.
This is crucial to describe exclusive processes involving, for instance, pions.

\section*{Acknowledgements}
I thank Matthias Berwein, Nora Brambilla, Abhishek Mohapatra and Tommaso Scirpa for collaboration on the work presented here.

\end{document}